\documentclass[prb,aps,amssymb,twocolumn,showpacs,floatfix]{revtex4}

\usepackage{graphicx}
\usepackage{dcolumn}
\usepackage{bm}
\usepackage{url}

\begin{document}

\title{Tunability of Superconducting Metamaterials}
\author{Michael C. Ricci}
\author{Hua Xu}
 \affiliation{Center for Superconductivity, Department of Physics, University of Maryland, College Park, MD 20742-4111}
\author{Ruslan Prozorov}
 \affiliation{Ames Laboratory, Department of Physics \& Astronomy, Iowa State University, Ames, IA 50011}
\author{Alexander P. Zhuravel}
 \affiliation{B. Verkin Institute for Low Temperature Physics \& Engineering, National Academy of Sciences of Ukraine, 61164 Kharkov, Ukraine}
\author{Alexey V. Ustinov}
 \affiliation{Physics Institute III, University of Erlangen-Nuremberg, D-91508, Erlangen, Germany}
\author{Steven M. Anlage}
 \affiliation{Center for Superconductivity, Department of Physics, University of Maryland, College Park, MD 20742-4111}
 
\date{\today}

\begin{abstract}
Metamaterials are artificial structures with unique electromagnetic properties, such as relative dielectric permittivity and magnetic permeability with values less than $1$, or even negative. Because these properties are so sensitive to loss, we have developed metamaterials comprised of superconducting waveguides, wires, and split-ring resonators. An important requirement for applications of these metamaterials is the ability to tune the frequency at which the unique electromagnetic response occurs. In this paper we present three methods (unique to superconductors) to accomplish this tuning: temperature, dc magnetic field, and rf magnetic field.  Data are shown for dc and rf magnetic field tuning of a single Nb split-ring resonator (SRR).  It was found that the dc field tuning was hysteretic in the resonant frequency data, while the quality factor, $Q$, was less hysteretic.  The rf power tuning showed no hysteresis, but did show supression of the $Q$ at high power.  Magneto-optical images reveal inhomogeneous magnetic vortex entry in the dc field tuning, and laser scanning photoresponse images for a YBa$_2$Cu$_3$O$_{7-\delta}$ SRR reveals the current distribution in the rings.
\end{abstract}

 \maketitle

	Objects with a negative index of refraction were first theorized in detail by Veselago in 1967 \cite{veselago:509}; however, such objects were unexplored experimentally until the late 1990s when the technology became available to realize their properties by means of artificial materials, initially at microwave frequencies \cite{pendry:37}.  Initial experiments utilized left-handed metamaterials (LHMs), composed of arrays of metallic wires, and split-ring resonators (SRRs), as explained in Section II.  Photonic crystals, and left-handed transmission lines (LHTLs) are also a focus of negative index of refraction research \cite{pendry:37}, and recently superconducting LHTLs have been explored \cite{salehi:996, wang:1}.
	
	A novel property of LHMs is the amplification of evanescent waves, which allows for an imaging resolution well below the Abbe limit ($\lambda /2$)  \cite{pendry:37}.  Normal metals used for the majority of LHM experiments are lossy at microwave frequencies, even when the LHM is constructed using very good conductors such as Cu.  These losses prevent the amplification of the high-$k$ vector (evanescent) components of the electromagnetic field, which carry the most detailed spatial information of an object \cite{shen:3286}.  One method of reducing losses in an LHM is to use superconducting metals in the fabrication of the LHM \cite{ricci:034102}.
	
		The effective index of refraction of a metamaterial may be written as $n_{\rm eff}(f) = \sqrt{\epsilon_{\rm eff}(f) \mu_{\rm eff}(f)}$, where $\epsilon_{\rm eff}(f)$ is the effective permittivity, and $\mu_{\rm eff}(f)$ is the effective permeability, and $f$ is the frequency.  Veselago's calculations showed that it was possible to obtain $n_{\rm eff}(f) < 0$ if the conditions of $\epsilon_{\rm eff}(f) < 0$ and $\mu_{\rm eff}(f) < 0$ were satisfied for some frequency bandwidth \cite{veselago:509}.  Generally, an array of metallic wires is used to obtain $\epsilon_{\rm eff}(f) < 0$.  (Since the data presented here focuses on SRRs, see Refs. \cite{pendry:37, ricci:034102} for details on wire arrays.)  SRRs are used to obtain a frequency bandwidth of $\mu_{\rm eff}(f) < 0$.  Originally, SRRs consisted of two concentric metallic strips, each in the shape of a square or circular ring, with one gap in each strip diametrically opposite to each other (see Fig. 1(a)).  The capacitive gaps introduce a resonance, and the SRR may be roughly modeled as a simple $LC$ resonator \cite{ricci:264102}. 
		
		An effective permeability can be written for an array of SRRs due to their resonant properties, even if the metal used in the design of the SRR is non-magnetic.  The effective permeability has the general form of $\mu_{\rm eff}(f) = 1 - F f^2/ (f^2 - f_0^2 + i \Gamma f)$, where $F$ is the filling fraction (the ratio of the area enclosed by the SRR, to the square of the lattice parameter of the SRR array), $f_0$ is the resonant frequency, and $\Gamma$ is the loss parameter \cite{pendry:37}.  Plotting $\mu_{\rm eff}(f)$ versus frequency reveals a limited frequency bandwidth for which $\mu_{\rm eff}(f) < 0$.  This $\mu_{\rm eff}(f) < 0$ band is dependent on $\Gamma$, and for high loss, such as that in a thin film in the normal state, the frequency bandwidth of $\mu_{\rm eff}(f) < 0$ vanishes \cite{ricci:034102}.  The frequency bandwidth of $\mu_{\rm eff}(f) < 0$ can be adjusted to cover a large range of frequencies, from microwave \cite{pendry:37} to optical \cite{grigorenko:335}, by changing the dimensions of the SRR.  The adjustment may also be finely tuned with temperature \cite{ricci:264102}, or with dc or rf magnetic fields (presented here), if superconducting metals are used in the fabrication of the SRR.   
		
	Superconducting LHMs provide many advantages over normal metal LHMs.  One advantage is the low-loss nature of the superconducting metals.  In order to amplify the evanescently decaying high-$k$ wavenumber excitations, the LHM losses must be as low as possible.  For instance, when imaging through a flat slab of thickness $d$ in the electrostatic limit, the imaginary part of the effective permittivity, which is related to the losses, must obey $\epsilon_{{\rm eff}, 2} \ll e^{-2k_x d}$, where $k_x$ is the lateral wavenumber \cite{shen:3286}.  Superconductors can achieve this restriction at microwave frequencies \cite{ricci:034102}.  
	
	Another benefit of using superconducting metals is the ability to tune the LHM resonant features by changing the environment of the LHM \cite{ricci:264102}.  As stated earlier, the equation for $\mu_{\rm eff}(f)$ contains a limited frequency bandwidth where $\mu_{\rm eff}(f) < 0$.  This frequency bandwidth may be tuned in frequency by changing the temperature \cite{ricci:264102}, by changing a dc magnetic field applied parallel to the applied transverse magnetic field, ${\bf B}$, shown in Fig. 1(a), or by increasing the rf power, $P_{\rm rf}$, of the applied electromagnetic signal.  Tuning has also been explored in superconducting LHTLs \cite{salehi:996}.  Normal metal SRR tuning has been performed by using varactors to couple the two rings of an SRR \cite{gil:1347}.
	
	Finally, superconducting LHMs can be miniaturized while still maintaining their low-loss properties.  For comparison, as the size of normal metal wires and SRRs are decreased, losses increase as $\rho/r^2$ and $\rho/t \ell$, respectively, where $\rho$ is the resistivity, $r$ is the wire radius, $\ell$ is the size of the SRR, and $t$ is the thickness of the SRR.  As the normal metal SRR dimensions decrease, for example, losses will increase and the frequency bandwidth of $\mu_{\rm eff}(f) < 0$ will vanish.  This does not happen with superconducting SRRs until the critical current density is reached \cite{ricci:264102}.

\begin{figure}
\includegraphics[width=2.36 in]{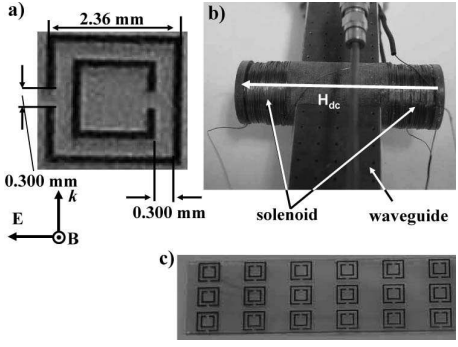}
\caption{Images of the materials: a) Nb (and YBa$_2$Cu$_3$O$_{7-\delta}$) SRR dimensions and the applied field geometry inside the Ag-plated Cu $X$-band waveguide; b) the solenoid straddling the Ag-plated Cu $X$-band waveguide used in the dc magnetic field experiment; c) Nb SRR array on a quartz chip.}
\end{figure}  	

	The experiments presented here were carried out in an Ag-plated Cu $X$-band waveguide with inner dimensions of $22.86 \times 10.16$ mm$^2$ containing a single Nb SRR.  The broadsides of the waveguide were perforated with $0.51$ mm diameter holes, spaced $5.08$ mm longitudinally and transversely, to allow for a wire array to be incorporated into the experiment.  The Nb SRR dimensions, and applied electromagnetic field geometry inside the Ag-plated Cu $X$-band waveguide, are shown in Fig. 1(a).  The Nb SRR was made by rf sputtering $200$ nm of Nb onto a $350$-micron-thick quartz substrate, and using standard photolithography and plasma etching.  The quartz was diced into chips with three rows by nine columns of Nb SRRs (Fig. 1(c)), and one chip was diced further to obtain the single Nb SRR ($T_c = 8.65$ K by ac susceptibility).
	
\begin{figure}
\includegraphics[width=2.36 in]{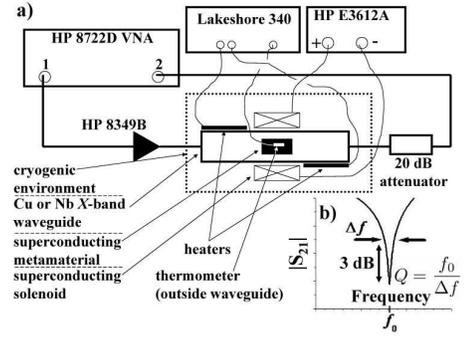}
\caption{a) Experimental setup.  The amplifier, solenoid with current source, and attenuator are all optional, and are not present at the same time. (They are all shown for completeness.)  b) shows an example of the type of transmission data ($S_{21}$) obtained in the experiment, and shows the definition of $Q$.}
\end{figure}	

Figure 2 shows the experimental setup.  The Ag-plated Cu $X$-band waveguide containing the single Nb SRR, centered in the waveguide, was submerged in liquid helium.  The sample was zero-field-cooled (ZFC) in a magnetically shielded cryostat, and the temperature was monitored by a Si diode, using a Lakeshore 340 Temperature Controller.  The stainless steel coated Cu antennas that injected the applied TE$_{10}$ field mode into the waveguide were coupled to an HP 8722D vector network analyzer (VNA), via $1.37$ m of Cu coaxial cable (type UT-85).  The VNA measured the $S$-matrix for the two ports, over $10$ and $200$ MHz spans about the center of the $|S_{21}|$ transmission dip ($\mu_{\rm eff}(f) < 0$ region) of the single Nb SRR.  See Fig. 2(b) for an example of the data.  The VNA was electronically calibrated to the outside of the cryostat over a frequency range of $0.5$ to $26.5$ GHz.

In the case of the applied dc magnetic field experiment, a solenoid straddled the Ag-plated Cu $X$-band waveguide (see Fig. 1(b)), and applied a dc magnetic field parallel to ${\bf B}$ in Fig. 1(a).  This solenoid was constructed by wrapping superconducting wire around a tube of Bakelite, and was powered by an HP E3612A DC current source.  The solenoid current was increased by hand, and measured with a multimeter.  The dc magnetic field of the solenoid was calibrated with a Guassmeter at room temperature. 

For the rf power experiments, an HP 8349B amplifier was coupled to Port 1 of the VNA, with the output of the amplifier coupling to the coaxial cables leading to the waveguide, and to avoid damaging the VNA (Fig. 2), a $20$ dB attenuator was placed before the second port of the VNA.  The VNA power was then incremented from $-30$ to $+5$ dBm.  The rf power applied to the sample, $P_{\rm rf}$, was assumed to be the VNA power, plus $20$ dB gain from the amplifier.  			
\begin{figure}
\includegraphics[width=2.36 in]{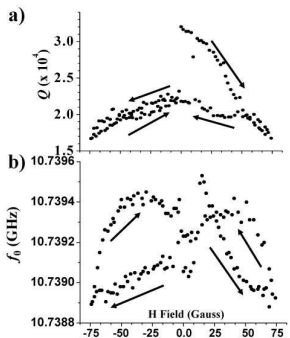}
\caption{Measurements of the single Nb SRR in an applied dc ${\bf H}$ field, at $4.2$ K.  Arrows indicate direction of field ramping.  Part a) shows the change in $Q$ and part b) shows the change in the resonant frequency with applied dc ${\bf H}$ field.}
\end{figure}

	Tuning of the frequency bandwidth of $\mu_{\rm eff}(f) < 0$ with temperature was shown in Ref. \cite{ricci:264102}.  Another method of tuning the frequency bandwidth of $\mu_{\rm eff}(f) < 0$ of the Nb SRR involves modifying the inductance and loss in the Nb film with magnetic vortices.  A dc magnetic field parallel ${\bf B}$ in Fig. 1(a) was applied to a single ZFC Nb SRR inside an Ag-plated Cu $X$-band waveguide, at a fixed temperature of $4.2$ K.  The applied dc magnetic field introduced vortices first at the defects of the Nb film, and then at the perimeter of the square rings.  The data are shown in Fig. 3.  As the applied dc magnetic field was increased, more vortices were introduced, primarily at corners and defects, and the resonant frequency, $f_0({\bf H})$, initially increased, then decreased.  The applied dc magnetic field was then decreased by decreasing the solenoid current.  Once the solenoid current reading was $0$ A, this current was reversed, and the process of increasing and decreasing the solenoid current was repeated.  A so-called butterfly pattern emerged in the data of $f_0({\bf H})$, shown in Fig. 3(b), where the arrows indicate the direction of the change of the applied dc magnetic field.   We speculate that the initial increase in $f_0({\bf H})$ at small fields, and the non-monotonic behavior, were due to the suppression of the edge and corner rf current build-up (which make a large contribution to the inductance) by the applied flux.
	
	The quality factor, $Q$, data are shown in Fig. 3(a), and the arrows show field ramping directions.  The results follow a similar trend as the frequency change (Fig. 3(b)), although the $Q$ data are essentially monotonic, and less hysteretic.  The initial value of $Q$ was $3.2 \times 10^4$, and the $Q$ never returns to this value for any value of the applied dc magnetic field. The hysteresis loop with a maximum $Q$ of $2.3 \times 10^4$ shows that the Nb SRR films enter a critical state.

	Magneto-optical (MO) imaging (Fig. 4(d)) was performed on the Nb SRR,  to image the mixed state of the Nb, and show how magnetic flux, which is applied in a similar manner to the dc magnetic field experiment described above, enters and leaves the metal \cite{jooss:651}.   The results are shown in Fig. 4.  The optical image, Fig. 4(a), clearly shows surface defects of the Nb film, such as a scratch on the left side of the outer ring.  In Fig. 4(b), a $75$ Oe flux was applied to the ZFC Nb SRR, at $5$ K.  The dark regions are where the Nb film was still in the Mei\ss ner state, whereas the brighter regions on the Nb film indicate where the applied flux was penetrating.  For this low field strength, the flux penetration came mostly from the defects in the film, and the scratch on the left side of the outer loop is clearly visible.  Finally, Fig. 4(c) shows the trapped flux in the SRR, at $4.8$ K.  In this image, the scratch on the left side of the outer square is once again visible, as well as some scratches on the top to the inner ring, because the flux can easily exit at these locations.  The complicated structure of Fig. 3(b) is likely due to the change in inductance of the SRR from the flux entry at particular defects in the rings.
	
\begin{figure}
\includegraphics[width=2.36 in]{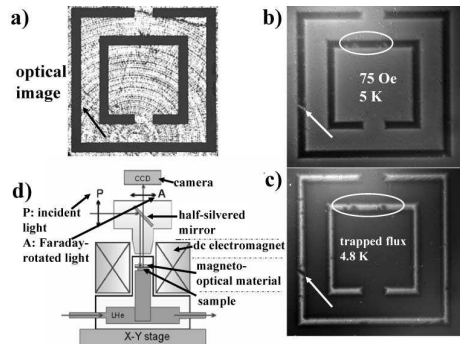}
\caption{Magneto-optical imaging of a single Nb SRR.  Part a) shows an optical image of the single Nb SRR.  Part b) shows the single ZFC Nb SRR in the presence of a $75$ Oe field, at $5$ K.  Part c) shows the trapped flux in the single Nb SRR, at $4.8$ K.  Part d) shows the experimental setup (see Ref. \cite{jooss:651}).  The arrows and circles in a) through c) indicate defects in the Nb film.}
\end{figure}
	
	An experiment was performed on the single ZFC Nb SRR in an Ag-plated Cu $X$-band waveguide, where rather than applying a dc magnetic field, the power of the incident signal was varied, at a fixed temperature of $4.2$ K and zero background field.  The $Q$ versus the rf power, $P_{\rm rf}$, data are shown in Fig. 5(a) and show a steady decrease in $Q$ with increasing power up to $+12$ dBm of incident rf power, at which point the $Q$ drops dramatically.  Figure 5(b) shows the resonant frequency, $f_0(P_{\rm rf})$, versus the rf power.  These results show that the resonant frequency did not respond to an increase in rf power, until about $+15$ dBm, at which point it suddenly increased by about $1$ MHz.  After this increase, the resonant frequency did not respond to increasing power, up to the amplifier output limit of +25 dBm.  Both the $Q$ and $f_0(P_{\rm rf})$ data showed little or no hysteresis as the rf power was increased and then decreased.	
	
	In the case of the rf power variations, there are high current densities that flow along the perimeter of the Nb SRR.  These currents build up on the edges and the inside corners of the Nb SRR, and can locally degrade the superconductivity of the Nb film.  Rather than observing the currents in the Nb SRR, a YBa$_2$Cu$_3$O$_{7-\delta}$ SRR was made, with the same dimensions as the Nb SRR, and a laser scanning photoresponse experiment was performed on the single YBa$_2$Cu$_3$O$_{7-\delta}$ SRR \cite{zhuravel:4979}.  The experimental setup is shown in Fig. 6(d).  The YBa$_2$Cu$_3$O$_{7-\delta}$ SRR was coupled to a YBa$_2$Cu$_3$O$_{7-\delta}$ microstrip, which carried the microwave signal.  The YBa$_2$Cu$_3$O$_{7-\delta}$ SRR, while held at a temperature of $80$ K, was illuminated with a laser, focused to a spot diameter of $\approx 1 \mu$m.  This illumination locally heated the YBa$_2$Cu$_3$O$_{7-\delta}$ film, and a change in the transmission parameter $|S_{21}|^2$ was measured as a function of the laser spot location.  The square root of this photoresponse, labeled $J_{\rm RF'}$, is proportional to the current distribution inside the YBa$_2$Cu$_3$O$_{7-\delta}$ rings \cite{zhuravel:4979}.
		
\begin{figure}
\includegraphics[width=2.36 in]{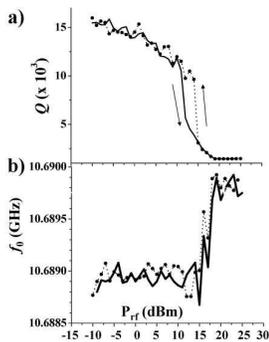}
\caption{Measurements of the rf power variations on a single Nb SRR, at $4.2$ K.  Part a) shows the change in $Q$ with the change in rf power.  Part b) shows the change in resonant frequency with the change in rf power.  The solid lines are increasing power, and the dotted lines are decreasing power.}
\end{figure}

\begin{figure}
\includegraphics[width=2.36 in]{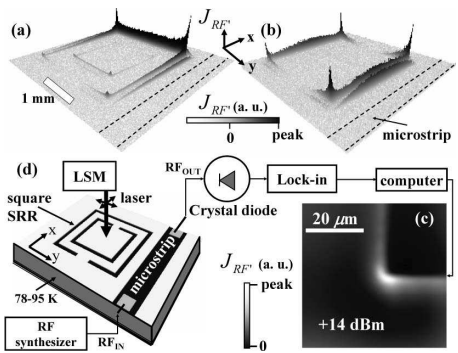}
\caption{Laser microscope photoresponse imaging for a single YBa$_2$Cu$_3$O$_{7-\delta}$ SRR at $80$ K.  Part a) shows $J_{\rm RF'}$ \cite{zhuravel:4979} for the fundamental mode; b) shows the second harmonic.  Part c) shows $J_{\rm RF'}$ for a closeup of an inside corner of the inner ring, at 9.57 GHz, and part d) shows the experimental setup.}
\end{figure}
	
	Figure 6(a) shows $J_{\rm RF'}$ for the fundamental mode ($5.146$ GHz) for both rings of the single YBa$_2$Cu$_3$O$_{7-\delta}$ SRR, and indicates the build-up of $J_{\rm RF'}$ in the inside corners.  Figure 6(b) shows the second harmonic ($10.838$ GHz), and the full-wave resonance was observed in the outer ring, while no significant $J_{\rm RF'}$ was observed in the inner ring.  It is noted again that there was a large current buildup in the corners of the YBa$_2$Cu$_3$O$_{7-\delta}$ rings.  A corner of the inner ring of the single SRR, with an applied power of $+14$ dBm and at a frequency of $9.57$ GHz, is shown in detail in Fig. 6(c).  The most likely explanation for the data in Fig. 5 is that as the rf power was increased, the current densities continued to build up at the corners (Fig. 6(c)) until a significant resistive component was generated \cite{zhuravel:1}. At some point this triggered runaway heating and severe degradation of the resonator $Q$, along with a reduction in inductance associated with the elimination of sharp-corner current buildup, and an increase in resonant frequency.

	It was shown that the frequency bandwidth of $\mu_{\rm eff}(f) < 0$ of superconducting LHMs was tunable in frequency while remaining at a fixed temperature, by changing a dc magnetic field applied parallel to the applied transverse magnetic field.  Increasing the rf power injected in the waveguide shifted the frequency bandwidth of $\mu_{\rm eff}(f) < 0$ in frequency as well, and showed supression of the $Q$ at high values of $P_{\rm rf}$.  Through MO imaging, vortices were found to come mostly from defects in the Nb film, and laser scanning photoresponse measurements indicated the importance of large currents that built up on the inside corners of the YBa$_2$Cu$_3$O$_{7-\delta}$ SRR rings.

The authors would like to thank N. Orloff for his help in the experiment.  This work was supported by the NSF grant No. ECS-0322844, and NASU grant ``Nanosystems, nanomaterials and nanotechnologies," and the German Science Foundation (DFG).


 \end{document}